\documentclass[aps,preprint]{revtex4}%
\usepackage{amsfonts}
\usepackage{amsmath}
\usepackage{amssymb}
\usepackage{graphicx}%
\begin{document}
\title{{\LARGE A GENERAL RELATIVISTIC ROTATING EVOLUTIONARY UNIVERSE }}
\author{Marcelo Samuel Berman$^{1}$}
\affiliation{$^{1}$Instituto Albert Einstein\ / Latinamerica
\ \ \ \ \ \ \ \ \ \ \ \ \ \ \ \ \ \ \ \ \ \ \ \ \ \ \ \ \ \ \ \ \ \ \ \ \ \ \ \ \ \ \ -
Av. Candido Hartmann, 575 - \ \# 17}
\affiliation{80730-440 - Curitiba - PR - Brazil email: msberman@alberteinsteininstitute.org}
\date{(original: 10 January, 2007; revised: 24 February, 2008).}

\begin{abstract}
We show that when we work with coordinate cosmic time, which is not proper
time, Robertson-Walker's metric, includes a possible rotational state of the
Universe. An exact formula for the angular speed and the temporal metric
coefficient, is found.

\end{abstract}
\maketitle

{\LARGE A GENERAL RELATIVISTIC ROTATING \ \ \ EVOLUTIONARY UNIVERSE}

\ \ {\large \ \ MARCELO SAMUEL BERMAN }

\bigskip

\bigskip Standard textbooks on Relativistic Cosmology, consider the usual
Robertson-Walker's metric, which would represent a homogeneous and isotropic
expanding Universe. However, the absence of rotation in the model, cast doubts
on its validity as representing a real Universe: it would require a rigorous
fine tuning, in order to keep, since inception, a non-rotating Universe. Since
the Universe has been observed to be expanding with acceleration, the presence
of a positive cosmological constant (\ $\Lambda$\ ) , has been stressed in the
last few years.

\bigskip

As a boundary condition, to be satisfied by Einstein's equations, the Machian
condition has been put forward, sometimes in disguise. Berman (2007; 2007a;
2007b) has suggested that the consideration of a zero-total-energy Universe,
might represent the Machian desired properties. Brans and Dicke (1961),
presented new field equations that would satisfy the approximate relation,

\bigskip

$\frac{GM}{c^{2}R}\sim1$ \ \ \ \ \ \ \ \ \ \ \ \ \ \ \ \ \ \ \ \ \ .

\bigskip

\bigskip In the above, \ \ $G$\ \ \ represents the gravitational constant, and
the\ \ mass \ \ $M$\ \ is that of a causally related Universe with radius
\ $R$\ \ \ .\ \ \ Sabbata and Sivaran (1994), have shown that a closely
related approximation would apply for the spin of the Universe \ $L$\ \ , namely,

\bigskip

$\frac{L^{2}}{c^{2}M^{2}R^{2}}\sim1$ \ \ \ \ \ \ \ \ \ \ \ \ \ \ .

\bigskip

We see that the Universe was expected to have a non-zero spin \ $L$\ \ .
Though both relations above are heuristic, Berman has shown in the cited
references, that the zero-total energy from the Newtonian point of view, could
yield several exact relations which substitute the above two approximations.
In conclusion: we expect a non-zero spin of the Machian Universe.

\bigskip

As we shall show below, by a simple expedient, we can obtain a rotational and
expanding model, out of the original Robertson-Walker's metric.

Gomide and Uehara (1981) derived the field equations for a Robertson-Walker's
metric in terms of coordinate time ( $t$\ ), when this time is not proper time
( $\tau$\ ). In the most simplest case, we may write:

\bigskip

$d\tau=(g_{00})^{1/2}dt$ \ \ \ \ \ \ \ \ \ \ \ \ \ \ \ \ \ \ \ \ \ \ \ \ , \ \ \ \ \ \ \ \ \ \ \ \ \ \ \ \ \ \ \ \ \ \ \ \ \ \ \ \ \ \ \ \ \ \ \ \ \ \ \ \ \ \ \ \ \ \ \ \ \ \ \ \ \ \ \ \ \ \ \ \ \ \ \ \ \ (1)

\bigskip

where,

\bigskip

$g_{00}=g_{00}(t)$ \ \ \ \ \ \ \ \ \ \ \ \ \ \ \ \ \ \ \ \ \ \ \ \ \ \ \ \ . \ \ \ \ \ \ \ \ \ \ \ \ \ \ \ \ \ \ \ \ \ \ \ \ \ \ \ \ \ \ \ \ \ \ \ \ \ \ \ \ \ \ \ \ \ \ \ \ \ \ \ \ \ \ \ \ \ \ \ \ \ \ \ \ \ \ \ \ (2)

\bigskip

The line element becomes:

\bigskip

$ds^{2}=-\frac{R^{2}(t)}{\left(  1+kr^{2}/4\right)  ^{2}}\left[  d\sigma
^{2}\right]  +g_{00}\left(  t\right)  $ $dt^{2}$
\ \ \ \ \ \ \ \ \ \ \ \ \ \ \ . \ \ \ \ \ \ \ \ \ \ \ \ \ \ \ \ \ \ \ \ \ \ \ \ \ \ \ \ \ \ \ \ \ \ \ \ \ \ \ \ \ \ \ \ \ \ (3)

\bigskip

The field equations, in \ General Relativity Theory (GRT) become:

\bigskip

$3\dot{R}^{2}=\frac{1}{3}\kappa(\rho+\frac{\Lambda}{\kappa})g_{00}%
R^{2}-3kg_{00}$ \ \ \ \ \ \ \ \ \ \ \ \ \ \ \ \ \ \ \ \ , \ \ \ \ \ \ \ \ \ \ \ \ \ \ \ \ \ \ \ \ \ \ \ \ \ \ \ \ \ \ \ \ \ \ \ \ \ \ \ \ \ \ \ \ \ \ \ (4)

\bigskip

and,

\bigskip

$6\ddot{R}=-g_{00}\kappa\left(  \rho+3p-2\frac{\Lambda}{\kappa}\right)
R-3g_{00}\dot{R}$ $\dot{g}^{00}$ \ \ \ \ \ \ \ \ \ \ \ . \ \ \ \ \ \ \ \ \ \ \ \ \ \ \ \ \ \ \ \ \ \ \ \ \ \ \ \ \ \ \ \ \ \ \ \ \ \ \ \ \ \ \ \ (5)

\bigskip

Local inertial processes are observed through proper time, so that the
four-force is given by:

\bigskip

$F^{\alpha}=\frac{d}{d\tau}\left(  mu^{\alpha}\right)  =mg^{00}$ $\ddot
{x}^{\alpha}-\frac{1}{2}m$ $\dot{x}^{\alpha}\left[  \frac{\dot{g}_{00}}%
{g_{00}^{2}}\right]  $ \ \ \ \ \ \ \ \ \ \ \ \ \ \ . \ \ \ \ \ \ \ \ \ \ \ \ \ \ \ \ \ \ \ \ \ \ \ \ \ \ \ \ \ \ \ \ \ \ \ \ \ \ (6)

\bigskip

Of course, when \ $g_{00}=1$\ \ , the above equations reproduce conventional
Robertson-Walker's field equations.

\bigskip

We must mention that the idea behind Robertson-Walker's metric is the Gaussian
coordinate system. Though the condition \ \ $g_{00}=1$\ \ is usually adopted,
we must remember that, the resulting time-coordinate is meant as representing
proper time. If we want to use another coordinate time, we still keep the
Gaussian coordinate properties.

\bigskip

From the energy-momentum\ \ conservation equation, in the case of a uniform
Universe, \ we must have,

\bigskip

$\frac{\partial}{\partial x^{i}}\left(  \rho\right)  =\frac{\partial}{\partial
x^{i}}\left(  p\right)  =\frac{\partial}{\partial x^{i}}\left(  g_{00}\right)
=0$ \ \ \ \ \ \ \ \ \ \ \ \ \ ( \ $i=1,2,3$\ \ ) \ \ \ \ . \ \ \ \ \ \ \ \ \ \ \ \ \ \ \ \ \ \ \ \ \ \ \ (7)

\bigskip

The above is necessary in the determination of cosmic time, for a commoving
observer. We can see that the hypothesis \ (2) -- that \ $g_{00}$\ \ is only
time-varying -- is now validated.

\bigskip

In order to understand equation (6)\ , it is convenient to relate the
rest-mass $m$\ , with \ an inertial mass \ $M_{i}$\ , with:

\bigskip

$M_{i}=\frac{m}{g_{00}}$\ \ \ \ \ \ \ \ \ \ \ \ \ \ .\ \ \ \ \ \ \ \ \ \ \ \ \ \ \ \ \ \ \ \ \ \ \ \ \ \ \ \ \ \ \ \ \ \ \ \ \ \ \ \ \ \ \ \ \ \ \ \ \ \ \ \ \ \ \ \ \ \ \ \ \ \ \ \ \ \ \ \ \ \ \ \ \ \ \ \ \ \ \ \ \ \ \ \ \ \ \ (8)

\bigskip

It can be seen that \ $M_{i}$\ \ represents the inertia of a particle, when
observed along cosmic time, i.e., coordinate time. In this case, we observe
that we have two acceleration terms, which we call,

\bigskip

\bigskip\ $a_{1}^{\alpha}=\ddot{x}^{\alpha}$\ \ \ \ \ \ \ \ \ \ \ \ \ \ \ , \ \ \ \ \ \ \ \ \ \ \ \ \ \ \ \ \ \ \ \ \ \ \ \ \ \ \ \ \ \ \ \ \ \ \ \ \ \ \ \ \ \ \ \ \ \ \ \ \ \ \ \ \ \ \ \ \ \ \ \ \ \ \ \ \ \ \ \ \ \ \ \ \ \ \ \ \ \ \ \ \ \ \ \ \ \ \ \ (9)

\bigskip

and,

\bigskip

\bigskip\ $a_{2}^{\alpha}=-\frac{1}{2g_{00}}\left(  \dot{x}^{\alpha}\dot
{g}_{00}\right)  $%
\ \ \ \ \ \ \ \ \ \ \ \ \ \ \ .\ \ \ \ \ \ \ \ \ \ \ \ \ \ \ \ \ \ \ \ \ \ \ \ \ \ \ \ \ \ \ \ \ \ \ \ \ \ \ \ \ \ \ \ \ \ \ \ \ \ \ \ \ \ \ \ \ \ \ \ \ \ \ \ \ \ \ \ \ \ \ \ (10)\bigskip

\bigskip

The first acceleration is linear; the second, resembles rotational motion.

\bigskip

If we consider \ \ $a_{2}^{\alpha}$\ \ a centripetal acceleration, we conclude
that the angular speed \ \ $\omega$\ \ \ is given by,

\bigskip

$\omega=\frac{1}{2}\left(  \frac{\dot{g}_{00}}{g_{00}}\right)  $%
\ \ \ \ \ \ \ \ \ \ \ \ \ \ \ \ \ . \ \ \ \ \ \ \ \ \ \ \ \ \ \ \ \ \ \ \ \ \ \ \ \ \ \ \ \ \ \ \ \ \ \ \ \ \ \ \ \ \ \ \ \ \ \ \ \ \ \ \ \ \ \ \ \ \ \ \ \ \ \ \ \ \ \ \ \ \ \ \ \ \ \ \ \ (11)

\bigskip

By comparison between the usual \ $\tau$ -- metric, and the field equations in
the \ $t$\ -- metric, we are led to conclude that the conventional energy
density \ $\rho$\ \ \ and cosmic pressure \ $p$\ \ \ are transformed into
\ \ $\bar{\rho}$\ \ \ and \ \ $\bar{p}$\ \ , where:

\bigskip

$\bar{\rho}=g_{00}\left(  \rho+\frac{\bar{\Lambda}}{\kappa}\right)
$\ \ \ \ \ \ \ \ \ \ \ \ \ \ \ \ , \ \ \ \ \ \ \ \ \ \ \ \ \ \ \ \ \ \ \ \ \ \ \ \ \ \ \ \ \ \ \ \ \ \ \ \ \ \ \ \ \ \ \ \ \ \ \ \ \ \ \ \ \ \ \ \ \ \ \ \ \ \ \ \ \ \ \ \ \ \ \ \ (12)

\bigskip

and,

\bigskip

$\bar{p}=g_{00}\left(  p-\frac{\bar{\Lambda}}{\kappa}\right)  $%
\ \ \ \ \ \ \ \ \ \ \ \ \ \ \ \ . \ \ \ \ \ \ \ \ \ \ \ \ \ \ \ \ \ \ \ \ \ \ \ \ \ \ \ \ \ \ \ \ \ \ \ \ \ \ \ \ \ \ \ \ \ \ \ \ \ \ \ \ \ \ \ \ \ \ \ \ \ \ \ \ \ \ \ \ \ \ \ \ (13)

\bigskip

We plug back into the field equations, and find,

\bigskip

$\bar{\Lambda}=\Lambda-\frac{3}{2\kappa}\left(  \frac{\dot{R}}{R}\right)
\dot{g}^{00}$ \ \ \ \ \ \ \ \ \ \ \ \ \ . \ \ \ \ \ \ \ \ \ \ \ \ \ \ \ \ \ \ \ \ \ \ \ \ \ \ \ \ \ \ \ \ \ \ \ \ \ \ \ \ \ \ \ \ \ \ \ \ \ \ \ \ \ \ \ \ \ \ \ \ \ \ \ \ \ \ \ \ \ (14)

\bigskip

For a time-varying angular speed, considering an arc $\phi$\ , so that,

\bigskip

$\omega(t)=\frac{d\phi}{dt}=\dot{\phi}$\ \ \ \ \ \ \ \ \ \ \ \ \ \ \ \ \ , \ \ \ \ \ \ \ \ \ \ \ \ \ \ \ \ \ \ \ \ \ \ \ \ \ \ \ \ \ \ \ \ \ \ \ \ \ \ \ \ \ \ \ \ \ \ \ \ \ \ \ \ \ \ \ \ \ \ \ \ \ \ \ \ \ \ \ \ \ \ \ \ \ \ \ (15)

\bigskip

we find, from (11),

\bigskip

$g_{00}=Ce^{2\phi(t)}$ \ \ \ \ \ \ \ . \ \ \ ( \ $C$\ \ = constant )\ \ \ \ \ \ \ \ \ \ \ \ \ \ \ \ \ \ \ \ \ \ \ \ \ \ \ \ \ \ \ \ \ \ \ \ \ \ \ \ \ \ \ \ \ \ \ \ \ \ \ \ \ \ \ \ \ \ (16)

\bigskip

Returning to (14), we find,

\bigskip

$\bar{\Lambda}=\Lambda+\frac{3}{\kappa C}\left(  \frac{\dot{R}}{R}\right)
\omega e^{-2\phi(t)}$ \ \ \ \ \ \ \ \ \ \ \ \ \ . \ \ \ \ \ \ \ \ \ \ \ \ \ \ \ \ \ \ \ \ \ \ \ \ \ \ \ \ \ \ \ \ \ \ \ \ \ \ \ \ \ \ \ \ \ \ \ \ \ \ \ \ \ \ \ \ \ \ \ \ \ \ \ (17)

\bigskip

We see that there is no singularity in the above relations for \ $g_{00}%
$\ \ and \ \ $\bar{\Lambda}$\ \ .

\bigskip

We have found an exact singularity-free solution, in GRT, for a rotating
evolutionary Universe, derived from the original Robertson-Walker's metric,
endowed with a lambda-term. The second right-hand-side term in (17)\ \ can be
neglected as time passes by. \ The inertial mass \ $M_{i}$\ \ , depends on the
rotation of the expanding Universe, \ fulfilling Mach's Principle.

\bigskip

The \ $\tau$-- field equations\ \ represent the compass of inertia; the \ $t$
-- metric, represents the rotation relative to the first one. By solving the
field equations, one may find the cosmic pressure and energy density, for a
given equation of state.

\bigskip

G\"{o}del's\ \ Universe (Adler et al., 1975), in which the matter does not
uniquely determine the geometry, violates Mach's Principle, so that, there is
not a distinguished universal time-coordinate in such Universe. More than
that, the bulk matter of that Universe is commoving relative to a particular
reference frame, but this frame is not inertial. The compass of inertia should
rotate relative to the matter, or vice-versa, but the bulk matter represents
the "fixed starts", so that, according to Mach's Principle, both can not
rotate one relative to the other.

\bigskip

This is not the case in the present model. The background \ $\tau$\ -- metric
defines the "fixed stars" . The rotation becomes evident in the \ $t$\ --
metric. All observers are commoving. This is done in a singularity-free\ \ framework.

\bigskip

The rotation of the Universe is a subject dealt by Berman (2007, 2007a, 2007b).

\bigskip

\bigskip

\bigskip{\Large Acknowledgments}

\bigskip

\bigskip The chief editor of this journal, has contributed with suggestions,
towards the clarification of the material in this paper. To him, the author
expresses his recognition, and also to his intellectual mentors, now friends
and colleagues, M.M. Som and F.M. Gomide. He thanks the many other colleagues
that collaborate with him. The typing was made by Marcelo F. Guimar\~{a}es,
who I consider a friend and to whom my thanks are due for this and many other
collaborations. I am grateful for the support \ by Geni, Albert and Paula.

\bigskip\bigskip

{\Large References}

\bigskip

\bigskip Adler, R.J.; Bazin, M.; Schiffer, M. (1975) - \textit{Introduction to
General Relativity, }Second Edition, McGraw-Hill, New York.

Berman,M.S. (2007) - \textit{Introduction to General Relativity, and the
Cosmological Constant Problem}, Nova Science, New York.

Berman,M.S. (2007a) - \textit{Introduction to General Relativistic and
Scalar-Tensor Cosmologies}, Nova Science, New York.

Berman,M.S. (2007b) - \textit{The Pioneer Anomaly and a Machian Universe},
Astrophysics and Space Science, \textbf{312}, 275.

Brans, C.; Dicke, R.H. (1961) - Physical Review, \textbf{124,} 925.\ \ \ 

Gomide, F.M.; Uehara, M. (1981) - Astron. Astrophys., \textbf{95}, 362.

Sabbata, V. de; Sivaran, C. (1994) - \textit{Spin and Torsion in Gravitation,
}World Scientific, Singapore.
\end{document}